\documentclass[preprint2,longabstract]{aastex}

\slugcomment{To appear in The Astronomical Journal (AJ)}		

\shorttitle{THE ARAUCARIA PROJECT}
\shortauthors{Karczmarek et al.}

\begin{document}

\title{THE ARAUCARIA PROJECT: THE DISTANCE TO THE CARINA DWARF GALAXY FROM
INFRARED PHOTOMETRY OF RR~LYRAE STARS
\footnote{Based on data collected with the VLT/HAWK-I instrument at ESO Paranal Observatory, Chile, as a part of a programme 082.D-0123(B).}}

\author{Paulina Karczmarek}
\affil{Warsaw University Observatory, Al. Ujazdowskie 4, 00-478, Warsaw, Poland}
\email{pkarczmarek@astrouw.edu.pl}

\author{Grzegorz Pietrzy{\'n}ski}
\affil{Universidad de Concepci{\'o}n, Departamento de Astronomia, Casilla 160-C, Concepci{\'o}n, Chile}
\affil{Warsaw University Observatory, Al. Ujazdowskie 4, 00-478, Warsaw, Poland}
\email{pietrzyn@astrouw.edu.pl}

\author{Wolfgang Gieren}
\affil{Universidad de Concepci{\'o}n, Departamento de Astronomia, Casilla 160-C, Concepci{\'o}n, Chile}
\affil{Millenium Institute of Astrophysics, Santiago, Chile}
\email{wgieren@astro-udec.cl}

\author{Ksenia Suchomska}
\affil{Warsaw University Observatory, Al. Ujazdowskie 4, 00-478, Warsaw, Poland}
\email{ksenia@astrouw.edu.pl}

\author{Piotr Konorski}
\affil{Warsaw University Observatory, Al. Ujazdowskie 4, 00-478, Warsaw, Poland}
\email{piokon@astrouw.edu.pl}

\author{Marek G{\'o}rski}
\affil{Millenium Institute of Astrophysics, Santiago, Chile}
\affil{Universidad de Concepci{\'o}n, Departamento de Astronomia, Casilla 160-C, Concepci{\'o}n, Chile}
\email{mgorski@astrouw.edu.pl}

\author{Bogumi\l{} Pilecki}
\affil{Warsaw University Observatory, Al. Ujazdowskie 4, 00-478, Warsaw, Poland}
\affil{Universidad de Concepci{\'o}n, Departamento de Astronomia, Casilla 160-C, Concepci{\'o}n, Chile}
\email{pilecki@astrouw.edu.pl}

\author{Dariusz Graczyk}
\affil{Millenium Institute of Astrophysics, Santiago, Chile}
\affil{Universidad de Concepci{\'o}n, Departamento de Astronomia, Casilla 160-C, Concepci{\'o}n, Chile}
\email{darek@astro-udec.cl}

\and
\author{Piotr Wielg{\'o}rski}
\affil{Warsaw University Observatory, Al. Ujazdowskie 4, 00-478, Warsaw, Poland}
\email{pwielgorski@astrouw.edu.pl}

\begin{abstract}
We obtained single-phase near-infrared (NIR) magnitudes in the $J$- and $K$-band for a sample of 33 RR~Lyrae stars in the Carina dSph galaxy. Applying different theoretical and empirical calibrations of the NIR period-luminosity-metallicity relation for RR~Lyrae stars, we find consistent results and obtain a true, reddening-corrected distance modulus of $20.118\pm 0.017 \mbox{ (statistical)} \pm 0.11 \mbox{ (systematic)}$ mag. This value is in excellent agreement with the results obtained in the context of the Araucaria project from NIR photometry of Red Clump stars ($20.165 \pm 0.015$) and Tip of Red Giant Branch ($20.09 \pm 0.03 \pm 0.12$ mag in $J$-band, $20.14 \pm 0.04 \pm 0.14$ mag in $K$-band), as well as with most independent distance determinations to this galaxy. The near-infrared RRL method proved to be a reliable tool for accurate distance determination at the 5 percent level or better, particularly for galaxies and globular clusters that lack young standard candles, like Cepheids.
\end{abstract}

\section{Introduction}
The main goal of the Araucaria Project is to improve the calibration of the cosmic distance scale from accurate observations of various stellar distance indicators in nearby galaxies \citep[e.g.][]{gieren05, pietrzynski13eas}. In the long-term perspective this analysis is expected to lead to a detailed understanding of the impact of metallicity or/and age on the various standard candles that are fundamental to calibrate the first rugs of the cosmic distance ladder. Errors in distance measurement methods propagate up the distance ladder, so minimization of errors at the local scale can significantly improve the accuracy of secondary distance indicators, and thus improve the determination of the Hubble constant. Significant improvement in the accuracy of distance measurements is possible by employing near-infrared (NIR) photometry, which minimizes the influence of both interstellar extinction and metallicity on the brightness of most stellar distance indicators \citep[e.g.][]{pietrzynski03, gieren09}.

Several theoretical and empirical studies of RR~Lyrae (RRL) stars demonstrated that these stars are able to provide improved distances from their magnitudes in NIR domain, as compared to traditional optical studies \citep[e.g.][]{bono03}. \citet{longmore} were the first to show that RRL stars follow period-luminosity (PL) relation in the $K$-band. Important contribution were added by \citet{nemec}, who made a comprehensive analysis of IR properties of RRL stars, and \citet{bono01}, who put the first theoretical constrains on $K$-band PL relation of RRLs based on nonlinear convective pulsation models. 

The continuation of theoretical studies of RRL stars in NIR wavebands, carried out by \citet{bono03} and \citet{catelan}, yielded the period-luminosity-metallicity (PLZ) relation in NIR domain. They reported that the effect of metallicity on the luminosity of RRLs is noticeably smaller in IR domain as compared to the visual domain. Indeed, the metallicity term enters the PLZ relation with the coefficient of approximately 0.2~mag\,dex$^{-1}$ in $K$-band \citep{bono03}, while in $V$-band the same metallicity coefficient can reach 0.3~mag\,dex$^{-1}$ \citep{criscienzo}. Following theoretical studies, empirical PLZ relations reported by \citet{sollima08}, \citet{borissova}, and \citet{dekany} showed the metallicity coefficient to be even smaller, ranging from 0.05 to 0.17~mag\,dex$^{-1}$. While in the optical domain the effect of metallicity (and even more subtle factors, like [$\alpha$/Fe] ratio) on PLZ relation is extensively studied, truly precise calibration of the PLZ in the NIR requires further theoretical and empirical work, and is beyond the scope of this paper.
    
To sum up, detailed theoretical and empirical studies of RRL variables in NIR domain demonstrated a decrease in the scatter in PL plane as well as a decrease in the metallicity dependence towards longer wavelengths. These two NIR properties of RRL variables, together with the fact that the interstellar reddening has marginal effect on brightness of stars in the IR domain, make the near-infrared PLZ relation for RRL stars an excellent means for distance determination to galaxies with a numerous old stellar population. This successful technique has been already used to provide accurate distances to the LMC \citep{szewczyk08, borissova}, the SMC \citep{szewczyk09}, and the Sculptor dSph galaxy \citep{pietrzynski08}. In this paper, we report the first distance determination to the Carina dSph galaxy from an application of $J$- and $K$-band PLZ relations on the Carina RRL stars. This study complements the existing distance determinations to Carina (Table \ref{tab:cardist}) with a very competitive (accuracy at the 5\% level or better) new measurement.

\section{Observations, data reduction, and calibration}

\begin{figure}[!ht]
\includegraphics[width=\columnwidth]{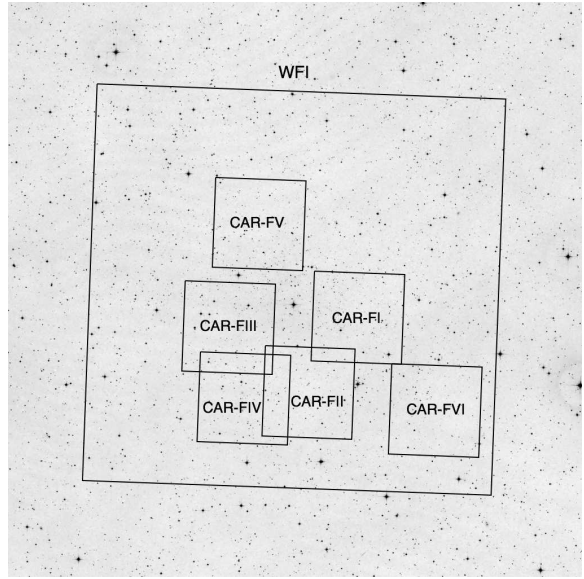}
\caption{
Six 7.5 $\times$ 7.5 arcmin VLT/HAWK-I fields in the Carina dSph galaxy, marked on the 50 $\times$ 50 arcmin DSS-2 infrared plate. The HAWK-I fields overlap the MPI/WFI field (34 $\times$ 33 arcmin) from \citet{dallora}. North is up and east is to the left.}
\label{fig:field}
\end{figure}

Near-infrared observations were conducted between 2008 November 11 and 2008 December 5 with the High Acuity Wide-field K-band Imager (HAWK-I) installed at the Nasmyth focus of UT4/VLT ESO 8\,m telescope at the Paranal observatory in Chile. The dates and coordinates of target fields are given in Table \ref{table:obs}. The location of six target fields in the Carina dSph galaxy was purposely chosen to overlap Wide Field Imager (WFI) field from \citet{dallora}, as shown in Figure \ref{fig:field}. The HAWK-I field of view of four 2048 $\times$ 2048 pixels detectors was about $7.5' \times 7.5'$ with a scale of 0.106'' pixel$^{-1}$ (detailed description of HAWK-I instrument available in \citet{kissler} and on the ESO website\footnote{http://www.eso.org/sci/facilities/paranal/instruments/hawki.html}). Six fields were observed in $J$ and $K_{\rm s}$ (further denoted as $K$) wavebands, under photometric conditions (seeing range $0.4-0.7"$). Our objects, having minimal brightness higher than 20 mag in $K$ and 21 mag in $J$, were well above the limiting HAWK-I magnitudes, i.e. 23.9 mag ($J$) and 22.3 mag ($K$) \citep{kissler}. In order to minimize the influence of sky background, the observations were made in a jitter mode, where the telescope was offset randomly between consecutive exposures, at a distance of 20'' from the original pointing.  Total integration times were 31 and 15 minutes for $K$- and $J$-band, respectively.

\subsection{HAWK-I reduction pipeline}
The HAWK-I reduction pipeline, a module of ESO Recipe Execution Tool (EsoRex), was utilized to execute complete reduction of our data. The reduction described below was executed simultaneously for all four chips of MEF (Multiple Extention FITS) file.
In the near-infrared domain the brightness of the atmosphere is significant, and therefore the background removal is crucial to calculate brightness of target objects. The reduction pipeline run this subtraction twice. The first background subtraction, using the median of the adjacent science images, took place after the basic calibration routines (dark correction, flat fielding and bad pixel correction). Before the second background subtraction, all frames were combined into one image, which was used in the object detection routine to derive the object mask. Only then the second background subtraction was performed for all frames, this time ignoring the pixels covered by the object mask. Each HAWK-I chip is affected in minor way (less than 0.3\% across the field) by distortions that originate from various causes: atmospheric refraction, non-planar surface of detectors. These errors were minimized by creating a distortion map consisting of a grid of x and y corrections shifts, and applying this map on each frame, in order to recover the undistorted position of the detected objects \citep{kissler}. Next, the images underwent the offset refinement procedure of sub-pixel accuracy, which abolished the shifts in the images caused by the jittering. As the final step, the images were combined again into one final image. Because images of standard stars had only one exposure frame, the procedure leading to the final product consisted of fewer steps: basic calibration (dark correction, flat fielding), sky subtraction, and distortion correction. For more details on data calibration, the reader is referred to the ESO website.

\subsection{Photometry and calibration}
The pipeline developed in the course of the Araucaria Project was used for photometry; in particular, the point-spread function (PSF) photometry and aperture corrections were applied using DAOPHOT and ALLSTAR programs in a way described by \citet{pietrzynski02}. In order to calibrate our photometric data to the standard system, 16 standard stars from the United Kingdom Infrared Telescope list \citep[UKIRT,][]{hawarden} were observed at different air masses and spanning a broad range in color, bracketing the colors of the RR~Lyrae stars in Carina. Taking into account the large number of standard stars, the accuracy of the zero point of our photometry was estimated to be about 0.02~mag. 

In order to make an external check of our photometric zero point, we performed two independent brightness comparisons using data collected with different telescopes. On the one hand, we filtered 143 brightest stars from our dataset, transformed them onto Two Micron All Sky Survey (2MASS) photometric system and then compared with 2MASS photometry. On the other hand, we used data collected under photometric conditions with the NTT/SOFI instrument at ESO La Silla Observatory, Chile\footnote{Programme 092.D-0295(B)}. The reduction procedure was analogous to the routine described above for HAWK-I data, only it was performed using IRAF packages in a way described by \citet*{pietrzynski02}. Point-spread function (PSF) photometry, including aperture corrections, was performed in the exact same way as for HAWK-I data. The calibration onto UKIRT photometric system was based on 10 standard stars, observed together with the target fields under the photometric conditions and at different air masses. The accuracy of photometry zero point of this calibration was estimated to be 0.02~mag. Finally, around 200 common stars from SOFI dataset and HAWK-I dataset, brighter than 17 mag, were paired and their calibrated photometric magnitudes were compared. This method allowed us to determine zero-point differences (in sense 2MASS/SOFI photometry minus our results), which are given in Table \ref{table:zp}. The results firmly support the accuracy of the absolute calibration. 

\subsection{Identification of variables}
\begin{figure}[!ht]
\includegraphics[width=\columnwidth]{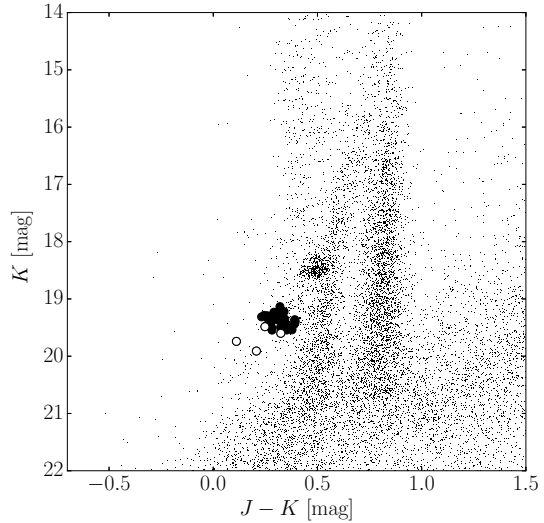}
\caption{
Infrared $J-K,~K$ CMD  showing identified RRab (black filled circles) and RRc (white filled circles) stars in observed fields. Although the Carina dSph galaxy is 22$^\circ$ from the galaxy plane, two pillars on the CMD plot which denote galactic disc dwarf stars ($J-K\approx 0.45$~mag) and halo giant stars ($J-K \approx 0.7$~mag) are clearly visible.}
\label{fig:cmd}
\end{figure}

The list of variables from \citet{dallora} served as a reference to cross-identify 33 RRL (29 RRab + 4 RRc) stars in our fields. Although our six HAWK-I fields overlapped WFI field in only 30\% (see Figure \ref{fig:field}), the sample of identified RRL stars constitutes 61\% of the sample of Dall'Ora. The position of our RRLs on the $K,J-K$ color-magnitude diagram (CMD) is shown in Figure \ref{fig:cmd}. All RRLs from our final sample have at least one measurement collected during photometric conditions. In case of 9 RRL stars, which were observed twice or found in two overlapping fields, we took a straight average of the random-phased magnitudes, which is expected to lead to a better approximation of their mean magnitudes. In case of single-measurement RRL stars, we took advantage of RRL infrared light curves (nearly sinusoidal and small-amplitude) and accepted single-phased measurements of $J$ and $K$ magnitudes as an approximations of reasonable accuracy to their mean magnitudes in these bands. Indeed, since RRL near-infrared amplitude is not larger than 0.4~mag \citep{marconi, szabo}, random-phased single measurements may cause deviation from the mean magnitude of maximum 0.2~mag for a single star. This error is extensively reduced by taking a large sample of stars. The $J$- and $K$-band random-phased magnitudes of the final RRL sample together with formal photometry errors from DAOPHOT and pulsational periods from the reference list \citep{dallora} are presented in Table \ref{tab:rrlyr}. In case of 9 RRL stars observed more than once, individual observations were put in separate rows.

\section{Distance determination}
In order to derive the apparent distance moduli to the Carina galaxy from our data, we used the following calibrations of the near-infrared PLZ relation for RRL stars: 

\begin{eqnarray}
M_K = -1.07 - 2.38 \log P + 0.08\,\mathrm{[Fe/H]_{CG}} \\
\mbox{\citep{sollima08}}& \nonumber
\end{eqnarray}

\begin{eqnarray}
M_K = -0.77 - 2.101 \log P + 0.231\,\mathrm{[Fe/H]_{CG}} \\
\mbox{\citep{bono03}} \nonumber
\end{eqnarray}

\begin{eqnarray}
M_K = -0.597 - 2.353 \log P + 0.175 \log Z \\
\mbox{\citep{catelan}} \nonumber
\end{eqnarray}

\begin{eqnarray}
M_J = -0.141 - 1.773 \log P + 0.190 \log Z \\
\mbox{\citep{catelan}} \nonumber
\end{eqnarray}

\begin{eqnarray}
M_K =  -0.6365 -2.347 \log P + 0.1747 \log Z \\
\mbox{\citep{dekany}} \nonumber
\end{eqnarray}

We recall that the calibration of \citet{sollima08} was constructed for the Two Micron All Sky Survey (2MASS) photometric system\footnote{~Note that in the following we will use $K$-band notation for $K_{\rm s}$ (2MASS)}, the calibrations of \citet{bono03} and \citet{catelan} are valid for the Bessel, Brett and Glass system (BBG), and \citet{dekany} calibrated his formula onto VISTA photometric system. Therefore, we transformed our own data, calibrated onto UKIRT \citep{hawarden}, to the BBG and 2MASS systems using the transformations presented by \citet{carpenter}, and to the VISTA system through the equations provided by CASU\footnote{~\url{http://casu.ast.cam.ac.uk/surveys-projects/vista/technical/photometricproperties/sky-brightness-variation/view}}. Above equations also account for first-overtone pulsators, once their period has been ``fundamentalized'' by adding a value of 0.127 to the logarithm of their periods.  

\begin{figure}[!ht]
\includegraphics[width=\columnwidth]{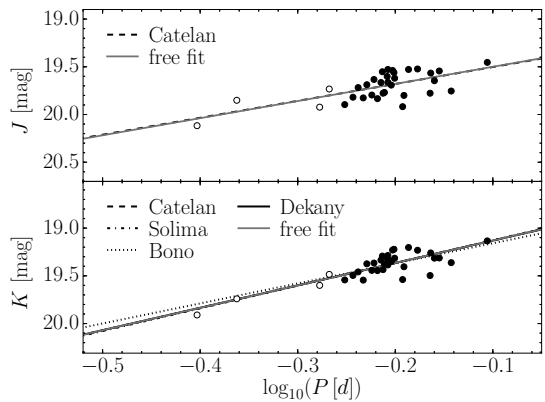}
\caption{
Period-luminosity relations for $J$- and $K$-bands defined by combined sample of RRL stars (RRab + ``fundamentalized'' RRc) observed in the Carina dSph galaxy, plotted along with the best fitted lines. The slopes of the fits were adopted from the recent theoretical and empirical calibrations, and the zero points were determined from our data. Filled and open circles represent RRab and RRc stars, respectively. The observed scatter is related mostly to the single phase nature of our photometry. Magnitude error bars are within the size of circles.}
\label{fig:zp}
\end{figure}

We calculated $K$- and $J$-band photometric zero point and slope for our sample of 33 RRLs using photometric magnitudes derived from our data, and pulsational periods excerpted from \citet{dallora}, as given in Table \ref{tab:rrlyr}. The linear least squares fitting to our data yielded the following slopes $-1.79 \pm 0.13$ and $-2.35 \pm 0.08$ in $J$- and $K$-band, respectively. We compared the obtained slopes with corresponding slopes from theoretical \citep{bono03, catelan} and empirical \citep{sollima08, dekany} PLZ relations (1)--(5). As seen in Figure \ref{fig:zp}, a large number of data points is concentrated within a small period range, which leads to an increase in uncertainties of both the slope and the zero point of our fit. In order to reduce this effect and gain accuracy of the zero point, we relied our further calculations on the slopes from formulae (1)--(5), provided that our slope values exhibit no virtual difference from the ``reference'' ones. We derived the apparent distance moduli by fitting the zero point for the fixed slopes of relations (1)--(5), assuming the mean metallicity of Carina $\rm{[Fe/H]=-1.72 \pm 0.25}$~dex \citep{koch06} in the \citet*{carretta} metallicity scale. 

In order to correct obtained apparent distance moduli for interstellar extinction, we adopted a foreground reddening calculated towards the Carina galaxy from the Galactic extinction maps of \citet{schlegel} along with the reddening law $R_V=3.1$ \citep{cardelli}. We assumed the value $E(B-V)=0.06$~mag to be an average of our six target fields (for discussion see section 4.2). We obtained the following values of extinction in near-infrared bands: $A_J=0.054$~mag, $A_K=0.022$~mag. The resulting true distance moduli for the Carina dSph galaxy with associated uncertainties are summarized in Table \ref{tab:truedist}. We adopt averaged value of $20.118\pm 0.017$ mag as the true distance modulus to the Carina galaxy. The systematic uncertainties are discussed in section~4.1.

\section{Discussion}
The distance moduli obtained from several independent theoretical and empirical calibrations of the infrared RRL period-luminosity-metallicity (PLZ) relations are consistent. The maximum difference of 0.06~mag between the results from the calibration of \citet{sollima08} and \citet{bono03} is not significant taking into account all uncertainties. Noteworthy, the similar discrepancy was recently reported for the SMC \citep{szewczyk09}, the LMC \citep{szewczyk08}, and the Sculptor galaxy \citep{pietrzynski08}. One of possible explanations is the underestimation of the effect of metallicity in Equation (1) relative to the rest of calibrations (2)--(5). The other possible explanation is the zero-point offset, in the sense that the distance modulus from the calibration of \citet{sollima08} is slightly shorter comparing to those from the calibration of \citet{bono03}. The earlier zero point calibration of PLZ relation of \citet{sollima06}, based on globular clusters, showed even larger offset (increased by 0.03 mag), and so we consider the present difference of 0.06~mag as an improvement in consistency between \citet{bono03} and \citet{sollima08}. 

A shorter distance modulus of 19.94 mag, calculated by \citet{udalski},  was tied to the underestimated LMC distance modulus (18.24 mag) derived by him from optical photometry of the Red Clump stars, and served as a point of reference for the distance determination to Carina. Providing a corrected distance modulus to the LMC of 18.493 mag \citep{pietrzynski13nat}, Udalski's result can be recalculated, yielding a value in close agreement with the present work.
 
\subsection{Uncertainties}
The uncertainties associated with our results are twofold. The statistical error (expressed as the standard deviation from the fit) contributes in values of 0.019~mag ($J$) and 0.015~mag ($K$) to the overall distance error, and accounts for: 
(i) the intrinsic scatter caused by single-phased measurements, 
(ii) the photometric error on each star, and
(iii) the uncertainty of the atmospheric extinction correction.
The systematic error is associated with: 
(i) the uncertainty of adopted mean metallicity,
(ii) the uncertainty of the photometric zero point,
(iii) the uncertainty of the reddening correction, and
(iv) the uncertainty of the zero point of the distance modulus fit.

The last source of the systematic error addresses the possibility that at the moment of observations a significant number of RRL stars from our sample could have brightnesses above (below) their mean luminosities. Since we treated single-phased magnitudes as mean magnitudes, these stars would bias the zero point of the distance modulus fit towards higher (lower) values. In order to evaluate this error, we generated a sample of 30 RRL light curves, and used them to calculate distances for two cases: i) the mean brightness is known from complete light curve, ii) the mean brightness is approximated by a single-phased measurement, i.e. a single point taken at random from each light curve. Any difference in distance calculated for these two cases was treated as a contribution to the overall bias. Repeating this procedure 1000 times for different amplitudes and shapes of light curves (tooth-saw, semicircle, and sinusoid), we calculated the mean offset between distances derived from complete light curves and from one random-phased measurements of $0.001 \pm 0.02$~mag. Therefore we adopt 0.02 mag as additional source of the systematic error. We recall that in case of stars with a couple of observation points, we took a straight average of random-phased magnitudes, which led to a better approximation of the mean magnitudes for these stars, and to a reduction in the uncertainty of the zero point fit.

The systematic error of a maximum value of 0.11 mag (see Table \ref{tab:truedist}) surpasses significantly the statistical error, and is dominated by the metallicity uncertainty. Taking into account errors from both sources, our best distance determination to the Carina dSph galaxy is $20.118\pm 0.017 \mbox{ (statistical)} \pm 0.11 \mbox{ (systematic)}$ mag.

\subsection{Extinction}
In the study, we adopted foreground reddening value of $E(B-V)=0.06$~mag, based on the Galactic extinction maps of \citet{schlegel}. We treated internal extinction in Carina as insignificant due to the fact, that low-mass, old and evolved objects, such as RRL stars, are very unlikely to be significantly obscured by the interstellar gas from dynamic evolution or mass outflows. Because many authors chose the foreground reddening value of $E(B-V)=0.03$~mag \citep[e.g.][]{dallora, coppola, smecker94, monelli03}, we tested this value with our data, and received distance moduli changed only by 0.01~mag compared to the distances given in Table \ref{tab:truedist}. Therefore we claim the reddening $E(B-V)$ in range $0.03-0.06$ mag has insignificant effect (within the errors) on the distance derived from NIR photometry of RR~Lyrae stars. 

\subsection{Metallicity}
Because RRL stars in Carina are too faint to measure spectroscopically their metallicities, the most certain information on the metal abundance in Carina comes from spectroscopic measurements of much brighter Red Giant Branch (RGB) stars. In our work, we adopted a mean metallicity ${\rm [Fe/H]}=  -1.72$ dex, derived by \citet{koch06} from a large sample of 437 spectra of mixed-age population RGB stars. Old RR~Lyrae stars may have metallicities slightly below this average, but this issue is addressed in the adopted uncertainty $\sigma_{\rm [Fe/H]} = 0.25$ dex. Our result closely agrees with other RRL distance determinations at similar metallicities; \citet{dallora} found distance moduli $20.10 \pm 0.12$ mag for ${\rm [Fe/H]}= -1.7$ dex, and \citet{coppola} $20.09 \pm 0.07 \pm 0.10$ mag for ${\rm [Fe/H]}= -1.79$ dex. Slightly larger value of $20.18 \pm 0.05$ mag was found by \citet{namara11} for ${\rm [Fe/H]}= -2.0$ dex, yet still this determination is consistent (within the errors) with the present paper.

\section{Summary and conclusions}
We determined the distance to the Local Group Carina dwarf galaxy from single-phase near-infrared observations in $J$- and $K$-band of 33 RR~Lyrae stars. The distance moduls to Carina was calculated using different theoretical and empirical calibrations of the near-infrared PLZ relation for RRL stars. We advocate an averaged value $20.118\pm 0.017 \mbox{ (statistical)} \pm 0.11 \mbox{ (systematic)}$ mag as a true, reddening-corrected distance modulus to the Carina galaxy. Our results are consistent and agree with other distance determinations obtained from a number of different and independent techniques and in different wavebands, especially with Red Clump stars and Tip of Red Giant Branch examined in near-infrared domain, in the course of Araucaria Project \citep{pietrzynski03,pietrzynski09}. The method presented in this paper was successfully used to measure distances to four members of the  Local Group (LMC, SMC, Sculptor, and Carina), and was proved to be an excellent tool for accurate distance determination, particularly for galaxies and clusters which lack young standard candles, like Cepheids. Distances to the LMC and the SMC, derived from near-infrared PLZ relation for RRLs, are in excellent agreement with the results obtained from late-type eclipsing binaries \citep{pietrzynski13nat, graczyk14}. The main asset of using eclipsing binaries for distance determination is that the distance is determined in almost geometrical way, and thus is very weakly affected by population effects. The consistency of results from these two methods -- eclipsing binaries and near-infrared RRL variables -- proves the reliability of the latter. The near-infrared PLZ relation for RRL stars is a robust method for accurate distance measurements, being only little influenced by metallicity and reddening within a fair range of values. This study complements existing distance measurements to the Carina galaxy with a very competitive (accuracy at the 5\% level or better) new determination. 

\acknowledgments
We greatly appreciate constructive remarks from an anonymous referee that helped to improve this paper. We thank the staff of the ESO Paranal and La Silla observatories for their support during the observations. We also gratefully acknowledge financial support for this work from the Polish National Science Centre grant PRELUDIUM 2012/07/N/ST9/04246 and the TEAM subsidy from the Foundation for Polish Science (FNP). W.G., G.P. and D.G gratefully acknowledge financial support for this work from the BASAL Centro de Astrofisica y Tecnologias Afines (CATA) PFB-06/2007. W.G., M.G. and D.G. also gratefully acknowledge support from the Chilean Ministry of Economy, Development and Tourism's Millennium Science Initiative through grant IC 120009 awarded to the Millennium Institute of Astrophysics (MAS). This research has made use of the 2MASS Database.

\bibliography{Karczmarek_AJ_arxiv}

\begin{thebibliography}{}
\expandafter\ifx\csname natexlab\endcsname\relax\def\natexlab#1{#1}\fi

\bibitem[{{Bono} {et~al.}(2001){Bono}, {Caputo}, {Castellani}, {Marconi}, \&
  {Storm}}]{bono01}
{Bono}, G., {Caputo}, F., {Castellani}, V., {Marconi}, M., \& {Storm}, J. 2001,
  \mnras, 326, 1183

\bibitem[{{Bono} {et~al.}(2003){Bono}, {Caputo}, {Castellani}, {Marconi},
  {Storm}, \& {Degl'Innocenti}}]{bono03}
{Bono}, G., {Caputo}, F., {Castellani}, V., {et~al.} 2003, \mnras, 344, 1097

\bibitem[{{Borissova} {et~al.}(2009){Borissova}, {Rejkuba}, {Minniti},
  {Catelan}, \& {Ivanov}}]{borissova}
{Borissova}, J., {Rejkuba}, M., {Minniti}, D., {Catelan}, M., \& {Ivanov},
  V.~D. 2009, \aap, 502, 505

\bibitem[{{Cardelli} {et~al.}(1989){Cardelli}, {Clayton}, \&
  {Mathis}}]{cardelli}
{Cardelli}, J.~A., {Clayton}, G.~C., \& {Mathis}, J.~S. 1989, \apj, 345, 245

\bibitem[{{Carpenter}(2001)}]{carpenter}
{Carpenter}, J.~M. 2001, \aj, 121, 2851

\bibitem[{{Carretta} \& {Gratton}(1997)}]{carretta}
{Carretta}, E., \& {Gratton}, R.~G. 1997, \aaps, 121, 95

\bibitem[{{Catelan} {et~al.}(2004){Catelan}, {Pritzl}, \& {Smith}}]{catelan}
{Catelan}, M., {Pritzl}, B.~J., \& {Smith}, H.~A. 2004, \apjs, 154, 633

\bibitem[{{Coppola} {et~al.}(2013){Coppola}, {Stetson}, {Marconi}, {Bono},
  {Ripepi}, {Fabrizio}, {Dall'Ora}, {Musella}, {Buonanno}, {Ferraro},
  {Fiorentino}, {Iannicola}, {Monelli}, {Nonino}, {Pulone}, {Th{\'e}venin}, \&
  {Walker}}]{coppola}
{Coppola}, G., {Stetson}, P.~B., {Marconi}, M., {et~al.} 2013, \apj, 775, 6

\bibitem[{{Dall'Ora} {et~al.}(2003){Dall'Ora}, {Ripepi}, {Caputo},
  {Castellani}, {Bono}, {Smith}, {Brocato}, {Buonanno}, {Castellani}, {Corsi},
  {Marconi}, {Monelli}, {Nonino}, {Pulone}, \& {Walker}}]{dallora}
{Dall'Ora}, M., {Ripepi}, V., {Caputo}, F., {et~al.} 2003, \aj, 126, 197

\bibitem[{{D\'{e}k\'{a}ny} {et~al.}(2013){D\'{e}k\'{a}ny}, {Minniti},
  {Catelan}, {Zoccali}, {Saito}, {Hempel}, \& {Gonzalez}}]{dekany}
{D\'{e}k\'{a}ny}, I., {Minniti}, D., {Catelan}, M., {et~al.} 2013, \apjl, 776,
  L19

\bibitem[{{Di Criscienzo} {et~al.}(2004){Di Criscienzo}, {Marconi}, \&
  {Caputo}}]{criscienzo}
{Di Criscienzo}, M., {Marconi}, M., \& {Caputo}, F. 2004, \apj, 612, 1092

\bibitem[{{Gieren} {et~al.}(2005){Gieren}, {Pietrzy{\'n}ski}, {Soszy{\'n}ski},
  {Bresolin}, {Kudritzki}, {Minniti}, \& {Storm}}]{gieren05}
{Gieren}, W., {Pietrzy{\'n}ski}, G., {Soszy{\'n}ski}, I., {et~al.} 2005, \apj,
  628, 695

\bibitem[{{Gieren} {et~al.}(2009){Gieren}, {Pietrzy{\'n}ski}, {Soszy{\'n}ski},
  {Szewczyk}, {Bresolin}, {Kudritzki}, {Urbaneja}, {Storm}, {Minniti}, \&
  {Garc{\'{\i}}a-Varela}}]{gieren09}
---. 2009, \apj, 700, 1141

\bibitem[{{Girardi} \& {Salaris}(2001)}]{girardi}
{Girardi}, L., \& {Salaris}, M. 2001, \mnras, 323, 109

\bibitem[{{Graczyk} {et~al.}(2014){Graczyk}, {Pietrzy{\'n}ski}, {Thompson},
  {Gieren}, {Pilecki}, {Konorski}, {Udalski}, {Soszy{\'n}ski}, {Villanova},
  {G{\'o}rski}, {Suchomska}, {Karczmarek}, {Kudritzki}, {Bresolin}, \&
  {Gallenne}}]{graczyk14}
{Graczyk}, D., {Pietrzy{\'n}ski}, G., {Thompson}, I.~B., {et~al.} 2014, \apj,
  780, 59

\bibitem[{{Hawarden} {et~al.}(2001){Hawarden}, {Leggett}, {Letawsky},
  {Ballantyne}, \& {Casali}}]{hawarden}
{Hawarden}, T.~G., {Leggett}, S.~K., {Letawsky}, M.~B., {Ballantyne}, D.~R., \&
  {Casali}, M.~M. 2001, \mnras, 325, 563

\bibitem[{{Kissler-Patig} {et~al.}(2008){Kissler-Patig}, {Pirard}, {Casali},
  {Moorwood}, {Ageorges}, {Alves de Oliveira}, {Baksai}, {Bedin}, {Bendek},
  {Biereichel}, {Delabre}, {Dorn}, {Esteves}, {Finger}, {Gojak}, {Huster},
  {Jung}, {Kiekebush}, {Klein}, {Koch}, {Lizon}, {Mehrgan}, {Petr-Gotzens},
  {Pritchard}, {Selman}, \& {Stegmeier}}]{kissler}
{Kissler-Patig}, M., {Pirard}, J.-F., {Casali}, M., {et~al.} 2008, \aap, 491,
  941

\bibitem[{{Koch} {et~al.}(2006){Koch}, {Grebel}, {Wyse}, {Kleyna}, {Wilkinson},
  {Harbeck}, {Gilmore}, \& {Evans}}]{koch06}
{Koch}, A., {Grebel}, E.~K., {Wyse}, R.~F.~G., {et~al.} 2006, \aj, 131, 895

\bibitem[{{Longmore} {et~al.}(1986){Longmore}, {Fernley}, \&
  {Jameson}}]{longmore}
{Longmore}, A.~J., {Fernley}, J.~A., \& {Jameson}, R.~F. 1986, \mnras, 220, 279

\bibitem[{{Marconi} {et~al.}(2003){Marconi}, {Caputo}, {Di Criscienzo}, \&
  {Castellani}}]{marconi}
{Marconi}, M., {Caputo}, F., {Di Criscienzo}, M., \& {Castellani}, M. 2003,
  \apj, 596, 299

\bibitem[{{Mateo} {et~al.}(1998){Mateo}, {Hurley-Keller}, \& {Nemec}}]{mateo}
{Mateo}, M., {Hurley-Keller}, D., \& {Nemec}, J. 1998, \aj, 115, 1856

\bibitem[{{McNamara}(1995)}]{namara95}
{McNamara}, D.~H. 1995, \aj, 109, 1751

\bibitem[{{McNamara}(2011)}]{namara11}
---. 2011, \aj, 142, 110

\bibitem[{{Mighell}(1997)}]{mighell}
{Mighell}, K.~J. 1997, \aj, 114, 1458

\bibitem[{{Monelli} {et~al.}(2003){Monelli}, {Pulone}, {Corsi}, {Castellani},
  {Bono}, {Walker}, {Brocato}, {Buonanno}, {Caputo}, {Castellani}, {Dall'Ora},
  {Marconi}, {Nonino}, {Ripepi}, \& {Smith}}]{monelli03}
{Monelli}, M., {Pulone}, L., {Corsi}, C.~E., {et~al.} 2003, \aj, 126, 218

\bibitem[{{Nemec} {et~al.}(1994){Nemec}, {Nemec}, \& {Lutz}}]{nemec}
{Nemec}, J.~M., {Nemec}, A.~F.~L., \& {Lutz}, T.~E. 1994, \aj, 108, 222

\bibitem[{{Pietrzy{\'n}ski} \& {Gieren}(2002)}]{pietrzynski02}
{Pietrzy{\'n}ski}, G., \& {Gieren}, W. 2002, \aj, 124, 2633

\bibitem[{{Pietrzy{\'n}ski} {et~al.}(2003){Pietrzy{\'n}ski}, {Gieren}, \&
  {Udalski}}]{pietrzynski03}
{Pietrzy{\'n}ski}, G., {Gieren}, W., \& {Udalski}, A. 2003, \aj, 125, 2494

\bibitem[{{Pietrzy{\'n}ski} {et~al.}(2009){Pietrzy{\'n}ski}, {G{\'o}rski},
  {Gieren}, {Ivanov}, {Bresolin}, \& {Kudritzki}}]{pietrzynski09}
{Pietrzy{\'n}ski}, G., {G{\'o}rski}, M., {Gieren}, W., {et~al.} 2009, \aj, 138,
  459

\bibitem[{{Pietrzy{\'n}ski} {et~al.}(2008){Pietrzy{\'n}ski}, {Gieren},
  {Szewczyk}, {Walker}, {Rizzi}, {Bresolin}, {Kudritzki}, {Nalewajko}, {Storm},
  {Dall'Ora}, \& {Ivanov}}]{pietrzynski08}
{Pietrzy{\'n}ski}, G., {Gieren}, W., {Szewczyk}, O., {et~al.} 2008, \aj, 135,
  1993

\bibitem[{{Pietrzy{\'n}ski} {et~al.}(2013{\natexlab{a}}){Pietrzy{\'n}ski},
  {Graczyk}, {Gieren}, {Thompson}, {Pilecki}, {Udalski}, {Soszy{\'n}ski},
  {Koz{\l}owski}, {Konorski}, {Suchomska}, {Bono}, {Moroni}, {Villanova},
  {Nardetto}, {Bresolin}, {Kudritzki}, {Storm}, {Gallenne}, {Smolec},
  {Minniti}, {Kubiak}, {Szyma{\'n}ski}, {Poleski}, {Wyrzykowski}, {Ulaczyk},
  {Pietrukowicz}, {G{\'o}rski}, \& {Karczmarek}}]{pietrzynski13nat}
{Pietrzy{\'n}ski}, G., {Graczyk}, D., {Gieren}, W., {et~al.}
  2013{\natexlab{a}}, \nat, 495, 76

\bibitem[{{Pietrzy{\'n}ski} {et~al.}(2013{\natexlab{b}}){Pietrzy{\'n}ski},
  {Graczyk}, {Gieren}, {Thompson}, {Soszy{\'n}ski}, {Pilecki}, {Storm},
  {Konorski}, {Suchomska}, {Gallenne}, {Nardetto}, {Karczmarek}, \&
  {Gorski}}]{pietrzynski13eas}
---. 2013{\natexlab{b}}, EAS Publications Series, 64, 305

\bibitem[{{Schlegel} {et~al.}(1998){Schlegel}, {Finkbeiner}, \&
  {Davis}}]{schlegel}
{Schlegel}, D.~J., {Finkbeiner}, D.~P., \& {Davis}, M. 1998, \apj, 500, 525

\bibitem[{{Smecker-Hane} {et~al.}(1994){Smecker-Hane}, {Stetson}, {Hesser}, \&
  {Lehnert}}]{smecker94}
{Smecker-Hane}, T.~A., {Stetson}, P.~B., {Hesser}, J.~E., \& {Lehnert}, M.~D.
  1994, \aj, 108, 507

\bibitem[{{Sollima} {et~al.}(2008){Sollima}, {Cacciari}, {Arkharov},
  {Larionov}, {Gorshanov}, {Efimova}, \& {Piersimoni}}]{sollima08}
{Sollima}, A., {Cacciari}, C., {Arkharov}, A.~A.~H., {et~al.} 2008, \mnras,
  384, 1583

\bibitem[{{Sollima} {et~al.}(2006){Sollima}, {Cacciari}, \&
  {Valenti}}]{sollima06}
{Sollima}, A., {Cacciari}, C., \& {Valenti}, E. 2006, \mnras, 372, 1675

\bibitem[{{Szab{\'o}} {et~al.}(2014){Szab{\'o}}, {Ivezi{\'c}}, {Kiss},
  {Koll{\'a}th}, {Jones}, {Sesar}, {Becker}, {Davenport}, \& {Cutri}}]{szabo}
{Szab{\'o}}, R., {Ivezi{\'c}}, {\v Z}., {Kiss}, L.~L., {et~al.} 2014, \apj,
  780, 92

\bibitem[{{Szewczyk} {et~al.}(2009){Szewczyk}, {Pietrzy{\'n}ski}, {Gieren},
  {Ciechanowska}, {Bresolin}, \& {Kudritzki}}]{szewczyk09}
{Szewczyk}, O., {Pietrzy{\'n}ski}, G., {Gieren}, W., {et~al.} 2009, \aj, 138,
  1661

\bibitem[{{Szewczyk} {et~al.}(2008){Szewczyk}, {Pietrzy{\'n}ski}, {Gieren},
  {Storm}, {Walker}, {Rizzi}, {Kinemuchi}, {Bresolin}, {Kudritzki}, \&
  {Dall'Ora}}]{szewczyk08}
---. 2008, \aj, 136, 272

\bibitem[{{Udalski}(2000)}]{udalski}
{Udalski}, A. 2000, \actaa, 50, 279

\bibitem[{{Vivas} \& {Mateo}(2013)}]{vivas}
{Vivas}, A.~K., \& {Mateo}, M. 2013, \aj, 146, 141

\end{thebibliography}
\bibliographystyle{apj}


\onecolumn{
\begin{deluxetable}{lcccccc}
\tablewidth{0pc}
\tablecaption{
\label{tab:cardist}
Summary of recent distance determinations for the Carina dSph Galaxy, obtained with different stellar indicators, in the optical and near-infrared domains.}
\tablehead{
\colhead{$(m-M)_0$}				 		& \colhead{Method\tablenotemark{a}} 	& \colhead{Band} 	& \colhead{Reference}}
\startdata
$19.87 \phn \pm 0.11$ 					& CMD 	& $V$, $I$ 	& \citet{mighell} \\
$20.24 \phn \phd \pm $\nodata	& CMD	& $V$			& \citet{monelli03} \\
$19.96 \phn \pm 0.08$					& RC		& $I$			& \citet{udalski}\\
$19.96 \phn \pm 0.06$ 					& RC 	& $I$ 			& \citet{girardi} \\
$20.165 \pm 0.015$ 						& RC 	& $K$ 			& \citet{pietrzynski03} \\
$20.05 \phn \pm 0.09$ 					& TRGB & $I$ 			& \citet{smecker94} \\
$19.94 \phn \pm 0.08$					& TRGB &	$I$			& \citet{udalski}\\
$20.09 \phn \pm 0.03 \pm 0.12$ 	& TRGB & $J$ 			& \citet{pietrzynski09} \\
$20.14 \phn \pm 0.04 \pm 0.14$ 	& TRGB & $K$ 			& \citet{pietrzynski09} \\
$20.06 \phn \pm 0.12$ 					& DC 	& $V$ 			& \citet{mateo} \\
$20.17 \phn \pm 0.10$ 					& DC 	& $B$, $V$ 	& \citet{vivas} \\
$20.01 \phn \pm 0.05 $					& SXP	& $V$			& \citet{namara95} \\
$20.22 \phn \pm 0.05$					& DS		& $V$			& \citet{namara11} \\
$20.12 \phn \pm 0.08$ 					& HB/RRL 	& $I$ 	& \citet{smecker94} \\
$19.93 \phn \pm 0.08$					& RRL	& $V$			&	\citet{udalski}\\
$20.10 \phn \pm 0.12$ 					& RRL 	& $V$ 			& \citet{dallora} \\
$20.18 \phn \pm 0.05$					& RRL 	& $V$			& \citet{namara11}	\\
$20.09 \phn \pm 0.07 \pm 0.10$ 	& RRL 	& $B$, $V$ 	& \citet{coppola} \\
$20.118\pm 0.017 \pm 0.11$		& RRL	& $J$, $K$	& This paper \\
\enddata
\tablenotetext{a}{Distance determination based on: color-magnitude diagram (CMD), Red Clump stars (RC), Tip of Red Giant Branch stars (TRGB), dwarf Cepheids (DC), SX Phoenicis variables (SXP), $\delta$ Scuti variables (DS), Horizontal Branch stars (HB), RR~Lyrae variables (RRL).}
\end{deluxetable}
}

\begin{deluxetable}{ccccc}
\tablewidth{0pc}
\tablecaption{
\label{table:obs}
Observational information on the target fields.}
\tablehead{ \colhead{Field} & \colhead{Field}  & \colhead{R.A.}        & \colhead{Decl.}       & \colhead{Date of} \\
                   \colhead{No.}   & \colhead{name} & \colhead{(J2000.0)} & \colhead{(J2000.0)} & \colhead{observation} }
\startdata
1 & CAR-FI   & 06:41:06.16 & -51:01:09.7 & 2008-11-17 \\
2 & CAR-FII  & 06:41:32.33 & -51:07:22.8 & 2008-11-17, 2008-12-05\\
3 & CAR-FIII & 06:42:14.64 & -51:01:58.8 & 2008-11-20 \\
4 & CAR-FIV  & 06:42:06.59 & -51:07:53.2 & 2008-11-20 \\
5 & CAR-FV   & 06:41:58.41 & -50:53:22.2 & 2008-11-20 \\
6 & CAR-FVI  & 06:40:25.17 & -51:08:52.0 & 2008-11-17, 2008-12-05\\
\enddata
\end{deluxetable}

\begin{deluxetable}{ccc}
\tablewidth{0pc}
\tablecaption{
\label{table:zp}
Difference of zero point calibration between 2MASS and SOFI, and our data obtained with HAWK-I.}
\tablehead{ \colhead{Filter} & \colhead{2MASS -- HAWK-I}  & \colhead{SOFI -- HAWK-I}\\
\colhead{} & \colhead{(mag)}  & \colhead{(mag)}}
\startdata
$J$  & $0.019 \pm 0.049$       & $0.039 \pm 0.042$\\
$K$  & $0.036 \pm 0.071$       & $0.018 \pm 0.039$\\
\enddata
\end{deluxetable}

\onecolumn{
\begin{deluxetable}{ccccccccccc}
\rotate
\tablewidth{0pc}
\tablecaption{
\label{tab:rrlyr}
Random-phase $J$ and $K$ magnitudes of the complete sample of 33 RR Lyrae stars. In case of 9 RRL stars observed more than once, individual observations are placed in separate rows.}
\tablehead{ \colhead{Star ID\tablenotemark{1}} & \colhead{HAWK-I field} & \colhead{R.A.} & \colhead{Decl.} & \colhead{Period} & \colhead{Type} & \colhead{$J$} & \colhead{$\sigma_J$} & \colhead{$K$} & \colhead{$\sigma_K$} & \colhead{$J-K$} \\
\colhead{} & \colhead{} & \colhead{(J2000.0)} & \colhead{(J2000.0)} & \colhead{(d)} & \colhead{} &  \colhead{(mag)} & \colhead{(mag)} & \colhead{(mag)} & \colhead{(mag)} & \colhead{(mag)}}
\startdata
V30   & CAR-FIII	& 06:42:24.16 	& -51:02:39.1 & 0.619 	& ab & 19.601 & 0.010 & 19.318 	& 0.026 	& 0.286 \\
V31   & CAR-FIII 	& 06:42:22.98 	& -50:59:16.1 & 0.651 	& ab & 19.528 & 0.010 & 19.203 	& 0.019 	& 0.327 \\
V40   & CAR-FIV 	& 06:42:15.63 	& -51:06:59.9 & 0.394 	&  c & 19.923 	& 0.013 & 19.600 	& 0.021 	& 0.324 \\
V47   & CAR-FV   & 06:42:09.04 	& -50:53:53.4 & 0.324 	&  c & 19.851 	& 0.020 & 19.741 	& 0.028 	& 0.111 \\
V49   & CAR-FIII 	& 06:42:08.85 	& -51:01:11.8 & 0.686 	& ab & 19.566 & 0.012 & 19.260 	& 0.020 	& 0.308 \\
V57   & CAR-FV   & 06:42:02.80 	& -50:52:56.5 & 0.612 	& ab & 19.552 & 0.015 & 19.293 	& 0.022 	& 0.259 \\
V60   & CAR-FIV  	& 06:41:59.75 	& -51:06:38.2 & 0.615 	& ab & 19.769 & 0.014 & 19.377 	& 0.022 	& 0.392 \\
V65   & CAR-FV   & 06:41:55.77 	& -50:55:34.9 & 0.642 	& ab & 19.918 & 0.021 & 19.539 	& 0.025 	& 0.378 \\
V67   & CAR-FII  	& 06:41:52.66 	& -51:05:19.3 & 0.613 	& ab & 19.774 & 0.012 & 19.441 	& 0.022 	&  0.335 \\
V67   & CAR-FII  	& 06:41:52.66 	& -51:05:19.3 & 0.613 	& ab & 19.758 & 0.012 & 19.392 	& 0.022 	&  0.368 \\
V67   & CAR-FII  	& 06:41:52.66 	& -51:05:19.3 & 0.613 	& ab & 19.786 & 0.012 & 19.486 	& 0.023 	&  0.300 \\
V73   & CAR-FIV 	& 06:41:46.91 	& -51:07:06.9 & 0.571 	& ab & 19.819 & 0.014 & 19.496 	& 0.022 	& 0.323 \\
V74   & CAR-FV   & 06:41:43.79 	& -50:54:08.3 & 0.403 	&  c & 19.733 	& 0.017 & 19.485 	& 0.031 	& 0.248 \\
V77   & CAR-FII  	& 06:41:39.26 	& -51:05:39.5 & 0.605 	& ab & 19.797 & 0.014 & 19.437 	& 0.021 	& 0.361 \\
V77   & CAR-FII  	& 06:41:39.26 	& -51:05:39.5 & 0.605 	& ab & 19.871 & 0.014 & 19.454 	& 0.022 	& 0.417 \\
V85	  & CAR-FV  	& 06:41:35.66 	& -50:50:07.1 & 0.644 	& ab & 19.800 & 0.017 & 19.405 	& 0.025 	& 0.393 \\
V91	  & CAR-FII  	& 06:41:29.47 	& -51:04:28.4 & 0.720 	& ab & 19.754 & 0.010 & 19.339 	& 0.020 	& 0.415 \\
V91	  & CAR-FII  	& 06:41:29.47 	& -51:04:28.4 & 0.720 	& ab & 19.754 & 0.012 & 19.386 	& 0.018 	& 0.367 \\
V92   & CAR-FII  	& 06:41:28.77 	& -51:06:50.3 & 0.620 	& ab & 19.568 & 0.012 & 19.233 	& 0.022 	& 0.335 \\
V92	  & CAR-FII  	& 06:41:28.77 	& -51:06:50.3 & 0.620 	& ab & 19.758 & 0.013 & 19.539 	& 0.024 	& 0.220 \\
V105 & CAR-FII 	& 06:41:16.57 	& -51:09:02.8 	& 0.630 	& ab & 19.548 & 0.012 & 19.243 	& 0.020 	& 0.305 \\
V105 & CAR-FII  	& 06:41:16.57 	& -51:09:02.8 	& 0.630 	& ab & 19.693 & 0.012 & 19.388 	& 0.022 	& 0.308 \\
V116 & CAR-FI   	& 06:41:05.60 	& -51:00:28.2 	& 0.685 	& ab & 19.777 & 0.012 & 19.497 	& 0.023 	& 0.280 \\
V122 & CAR-FI   	& 06:40:56.95 	& -51:00:46.0 	& 0.627 	& ab & 19.539 & 0.010 & 19.230 	& 0.020 	& 0.309 \\
V125 & CAR-FI   	& 06:40:53.33 	& -50:58:53.4 	& 0.597 	& ab & 19.796 & 0.013 & 19.443 	& 0.023 	& 0.353 \\
V126 & CAR-FI   	& 06:40:52.31 	& -50:58:56.7 	& 0.560 	& ab & 19.896 & 0.014 & 19.543 	& 0.025 	& 0.353 \\
V127 & CAR-FI   	& 06:40:52.19 	& -50:59:20.9 	& 0.625 	& ab & 19.692 & 0.011 & 19.349 	& 0.020 	& 0.343 \\
V135 & CAR-VI  	& 06:40:46.44 	& -51:05:23.6 	& 0.590 	& ab & 19.575 & 0.011 & 19.266 	& 0.019 	& 0.311 \\
V135 & CAR-VI  	& 06:40:46.44 	& -51:05:23.6 	& 0.590 	& ab & 19.800 & 0.012 & 19.481 	& 0.021 	& 0.319 \\
V143 & CAR-VI  	& 06:40:38.25 	& -51:11:29.6 	& 0.601 	& ab & 19.665 & 0.013 & 19.335 	& 0.024 	& 0.330 \\
V159 & CAR-VI   	& 06:40:12.75 	& -51:09:35.6 	& 0.578 	& ab & 19.557 & 0.012 & 19.332 	& 0.020 	& 0.229 \\
V159 & CAR-VI  	& 06:40:12.75 	& -51:09:35.6 	& 0.578 	& ab & 19.879 & 0.014 & 19.586 	& 0.024 	& 0.293 \\
V179 & CAR-FV  	& 06:41:49.33 	& -50:54:10.6 	& 0.665 	& ab & 19.523 & 0.017 & 19.232 	& 0.024 	& 0.291 \\
V188 & CAR-VI  	& 06:40:12.37 	& -51:07:06.1 	& 0.600 	& ab & 19.621 & 0.012 & 19.358 	& 0.021 	& 0.263 \\
V188 & CAR-VI   	& 06:40:12.37 	& -51:07:06.1 	& 0.600 	& ab & 19.646 & 0.012 & 19.374 	& 0.021 	& 0.272 \\
V189 & CAR-VI   	& 06:40:09.97 	& -51:08:05.1 	& 0.700 	& ab & 19.544 & 0.010 & 19.313 	& 0.018 	& 0.232 \\
V197 & CAR-FI   	& 06:41:00.26 	& -51:02:39.7 	& 0.295 	&  c 	& 20.117 & 0.014 & 19.910 	& 0.032 	& 0.207 \\
V199 & CAR-FII  	& 06:41:19.65 	& -51:10:23.3 	& 0.784 	& ab	& 19.402 & 0.009 & 19.126 	& 0.018 	& 0.279 \\
V199 & CAR-FII  	& 06:41:19.65 	& -51:10:23.3 	& 0.784 	& ab	& 19.506 & 0.009 & 19.142 	& 0.018 	& 0.364 \\
V201 & CAR-FIII 	& 06:42:10.49 	& -50:59:07.5 & 0.692 	& ab & 19.646 & 0.010 & 19.315 	& 0.018 	& 0.332 \\
V202 & CAR-FIII 	& 06:42:22.61 	& -50:59:18.4 & 0.620 	& ab & 19.528 & 0.010 & 19.287 	& 0.021 	& 0.244 \\
V204 & CAR-FIII 	& 06:42:04.17 	& -51:01:51.3 & 0.629 	& ab & 19.560 & 0.018 & 19.222 	& 0.023 	& 0.340 \\
V206 & CAR-FIV  	& 06:42:04.21 	& -51:09:40.9 & 0.585 	& ab & 19.826 & 0.013 & 19.545 	& 0.025 	& 0.282 \\
\enddata
\tablenotetext{1}{Star IDs from \citet{dallora}}
\end{deluxetable}
}

\onecolumn{
\begin{deluxetable}{cccccc}
\tablewidth{0pc}
\tablecaption{
\label{tab:truedist}
True distance moduli determined from different calibrations.}
\tablehead{ \colhead{Filter} & \colhead{$K$} & \colhead{$K$} & \colhead{$K$} & \colhead{$K$} & \colhead{$J$}}
\startdata
Calibration 			& Sollima 			& Bono 			& D\'{e}k\'{a}ny & Catelan 		& Catelan \\
\hline
$(m-M)_0$			& 20.078 		& 20.142 		& 20.115 		& 20.116			& 20.140 \\
Statistical error	& \phn 0.016 	& \phn 0.016 	& \phn 0.016 	& \phn 0.016 	& \phn 0.019 \\
Systematic error 	& \phn 0.090	& \phn 0.110	& \phn 0.100	& \phn 0.100	& \phn 0.100 \\
\enddata
\tablecomments{Average foreground reddening $E(B-V ) = 0.06$~mag towards the Carina galaxy was calculated using reddening maps from \citet{schlegel}.}
\end{deluxetable}
}

\end{document}